\documentstyle[epsfig,12pt]{article}

\begin{document}
\setlength{\baselineskip}{5mm}

\noindent{\large\bf
\centerline{PRODUCTION MECHANISMS AND}  
\centerline{POLARIZATION OBSERVABLES FOR}
\centerline{$p+d\rightarrow \ ^3He+ \pi^o$ NEAR THRESHOLD}
}\vspace{4mm}

\centerline{
L. Canton $^{a,b}$, G. Pisent$^{a,b}$, 
W. Schadow$^{b,1}$, and J.P. Svenne$^{b,c}$
}


\vspace{2mm}

\centerline{\small
$^a$Istituto Nazionale di Fisica Nucleare, Padova, via Marzolo, n. 8 Italy }
\centerline{\small
$^b$Dipartimento di Fisica dell'Universit\`a, Padova, via Marzolo, n. 8 Italy}
\centerline{\small
$^c$Department of Physics and Astronomy, University of Manitoba, Winnipeg MB, 
Canada}
\centerline{\small
$^1$ Permanent Address: Capgemini Ernst \& Young, D\"usseldorf, Germany}

\vspace{2mm}

\begin{abstract}
Pion production at threshold from nucleon-deuteron 
collisions is considered,  with reference to the 
outgoing channel where the three-nucleon 
system is bound. 
The available experimental data are compared with 
calculations using accurate nuclear wavefunctions 
coming from rigorous solutions of the three-nucleon 
quantum mechanical equations. The dominant contributions for 
pion production are obtained through matrix elements 
involving pion-nucleon rescattering mechanisms in 
S- and P-waves. S-wave rescattering includes also
an isoscalar contribution which is enhanced 
because of the off-shell effects.  
P-wave rescattering includes also 
diagrams involving explicitly the $\Delta$ degrees of freedom.
It is found that the pion-nucleon S-wave off-shell effects
in the isospin-even channel are of considerable importance.
Initial-state interactions between the 
proton and the deuteron have in general minor effects on the 
spin-averaged and spin-dependent observables, except when
the polarization observable involves interference terms
amongst the various helicity amplitudes.

{\small {\bf Keywords:} Polarization phenomena.  
Pion production. Few-nucleon systems.}
\end{abstract}



\vspace{3mm}
\noindent{\large \bf  Introduction}
\vspace{2mm}

Since the seminal paper by Ruderman [1], these reactions 
have been often studied phenomenologically by means of the ``deuteron model'',
and still are nowadays [2]. On the contrary, we have followed a 
microscopic approach [3] based on explicit $\Delta$ excitation, 
two-body mechanisms with $\pi$N rescattering in S-wave, and 
supplemented by the one-body $\pi$NN vertex. 
Approaches of this type have been initiated in the late seventies 
[4-6] but since then not much progress has been
accomplished owing to the complexities connected with the treatment
of the $NN\leftrightarrow NN\pi$ inelasticities in the three-nucleon
system. Nowadays, the calculations can be performed with
more accurate knowledge of the nuclear wavefunctions, 
and by including a much larger number of intermediate three-nucleon 
states.
ISI (Initial State Interactions) can be calculated through an 
Alt-Grassberger-Sandhas scheme [7], where realistic nucleon-nucleon 
transition matrices are used. Anti-symmetrization prescriptions
due to identity of the nucleons can be fully taken into account.

We discuss here the results we have recently obtained by means of this
approach. Details about the model calculation are given in Refs. [3,8-10].

\vspace{1cm}
\vfill\eject

\vspace{4mm}
\noindent{\large \bf The model}
\vspace{4mm}

The starting point is the phenomenological 
low-energy interaction Lagrangian, coupling the pion with the
nucleon field
\begin{eqnarray}
\label{Lagrangiana}
{\cal L}_{\rm int}
& = &
{f_{\pi NN}\over m_\pi}
\bar\Psi\gamma^\mu\gamma^5\vec\tau\Psi \cdot \partial_\mu \vec \Phi 
-4\pi{\lambda^{}_I\over m_\pi^2}
\bar\Psi\gamma^\mu\vec\tau\Psi \cdot
\left[\vec\Phi \times \partial_\mu \vec \Phi\right]
-4\pi{\lambda^{}_O\over m_\pi}
\bar\Psi \Psi
\left[\vec\Phi \cdot \vec \Phi\right] \label{Lag-piN}
\end{eqnarray}
and with the $\Delta$ field
\begin{eqnarray}
\label{LagrangianaD}
{\cal L}^\Delta_{\rm int}= -
{f_{\pi N\Delta}\over m_\pi}
\left(\bar\Psi_\Delta^\mu\vec T\Psi \cdot \partial_\mu \vec \Phi
+ h.c.\right) \, .
\label{Lag-piD}
\end{eqnarray}
The calculations herein illustrated have been performed with the following set of
parameters: ${f_{\pi NN}^2 / 4 \pi} = 0.0735$, ${f_{\pi N\Delta}^2 / 4 \pi} = 0.32$,
$\lambda^{}_I = 0.045$ and $\lambda^{}_O = 0.006$. 
A crucial aspect is represented by the off-shell extrapolations of 
these parameters in the evaluation of the $\pi$-production matrix elements. 
For the S-waves terms we have ($t$ is the square of the transverse four-momentum
of the $\pi$N system)
\begin{equation}
\lambda^{OFF}_I = \lambda^{ON}_I {m_\rho^2 \over m_\rho^2 -t}
\ {\Lambda_\rho^2 \over \Lambda_\rho^2 -t} \ \ \ \ \ \ \ \ \ \ \ \ \ \ \ \ \ \ \ 
\lambda^{OFF}_O = \lambda^{ON}_O  {  
{a_{SR} + a_\sigma {m_\sigma^2 \over m_\sigma^2 -t}}
\over a_{SR} + a_\sigma}
\, .
\label{s-waves-off-shell}
\end{equation}
The form on the left denotes the isospin-odd contribution in terms of a 
$\rho$-exchange model, while on the r.h.s. we describe the isospin-even term 
as the combined effect of phenomenological short-range (SR) processes 
and an effective scalar-meson ($\sigma$) exchange. 
The two combined effects act in opposite directions [11]. The form on the 
right leads to an off-shell enhancement of the probability amplitude
for pion production in the scalar-isoscalar channel. It remains a still
open question if this form mimics other physical effects, 
such as short-range heavy-meson exchange
contributions [12]. 
Further details, such as the treatment of the fully antisymmetrized
matrix elements with respect to the Faddeev three-nucleon wavefunction,
or the treatment of ISI in the nucleon-deuteron channel, or finally
the description of the $\Delta$ propagation in the three-baryon system,
are not discussed here.  
For these aspects, reference is made to previous works [3,8-10].

\vspace{4mm}
\noindent{\large \bf Results}
\vspace{4mm}

\begin{figure}[hp]\centering
\epsfig{file=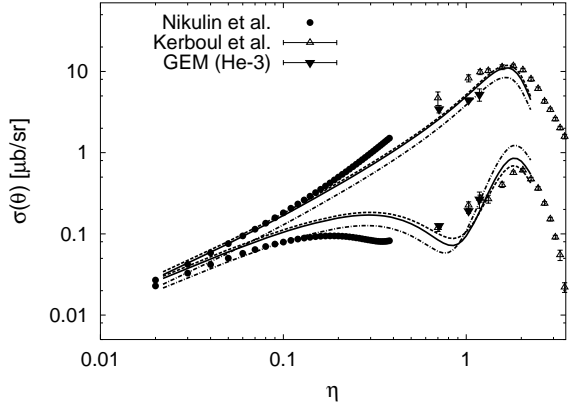, width=8cm, angle=-90}
\caption{Differential cross-sections for the 
$\vec{d}p\rightarrow {\ }^3He\ \pi^o$ reaction in collinear geometry. 
The parameter $\eta$ is the pion c.m. momentum in units of pion masses. 
Solid, dashed, dot-dashed lines refer, respectively, to 
results with Bonn B, Bonn A, and Paris potentials. Data from Ref.~[13-15].}
\end{figure}

\begin{figure}[hp]\centering
\epsfig{file=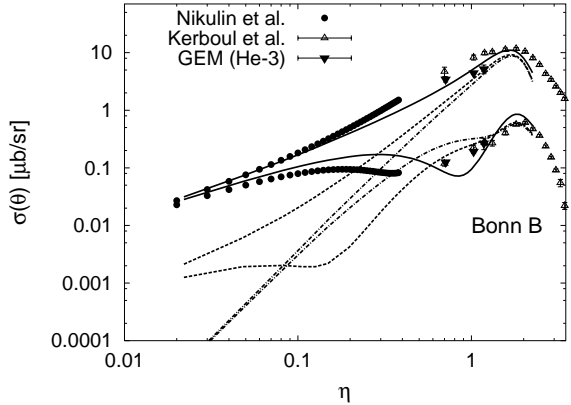, width=8cm, angle=-90}
\caption{Same as Fig.1. Calculation with the Bonn B potential.
Solid curve: full model; dashed line: setting to zero $\lambda_O$;
dot-dashed line: setting to zero also $\lambda_I$.}
\end{figure}

\begin{figure}[hp]\centering
\epsfig{file=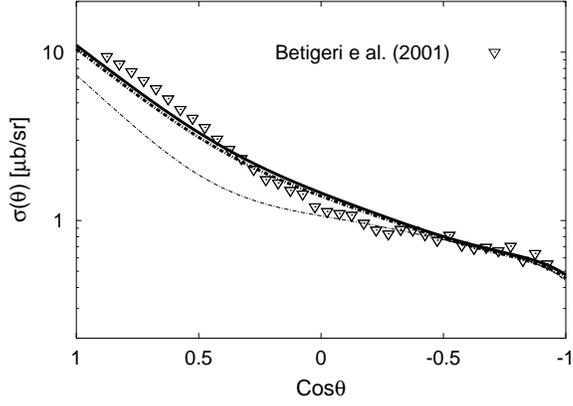, width=8cm, angle=-90}
\caption{Differential cross-section 
for the $pd\rightarrow \pi^+ \ {\ }^3H$ reaction
($850\ MeV/c$ proton beam). Results for the Paris potential.
Solid line: including ISI; dot-dashed line, without ISI;
thin line: setting to zero the isoscalar off-shell effects.
Data from Ref.~[15].}
\end{figure}

\begin{figure}[hp]\centering
\epsfig{file=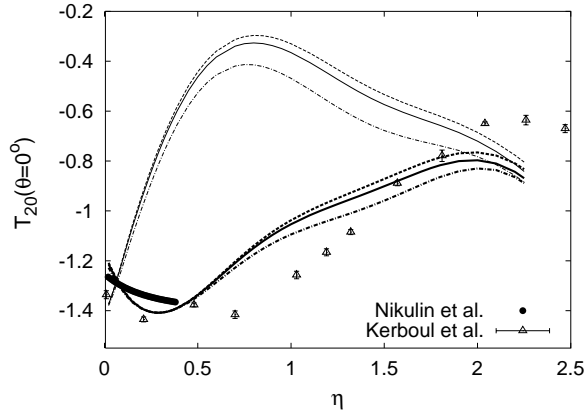, width=8cm, angle=-90}
\caption{Deuteron tensor analyzing powers in collinear geometry
(forward) for the $\vec{d}p\rightarrow {\ }^3He\ \pi^o$ reaction. 
The calculations are for different realistic NN potentials, following
the notation defined in Fig. 1.
Thin lines are obtained by setting to zero the isospin even S-wave contribution.
Thick lines are obtained by using the $\pi N$ model discussed in Ref. [11]. 
Data from Refs.~[13,14].}
\end{figure}

\begin{figure}
\centering
\epsfig{file=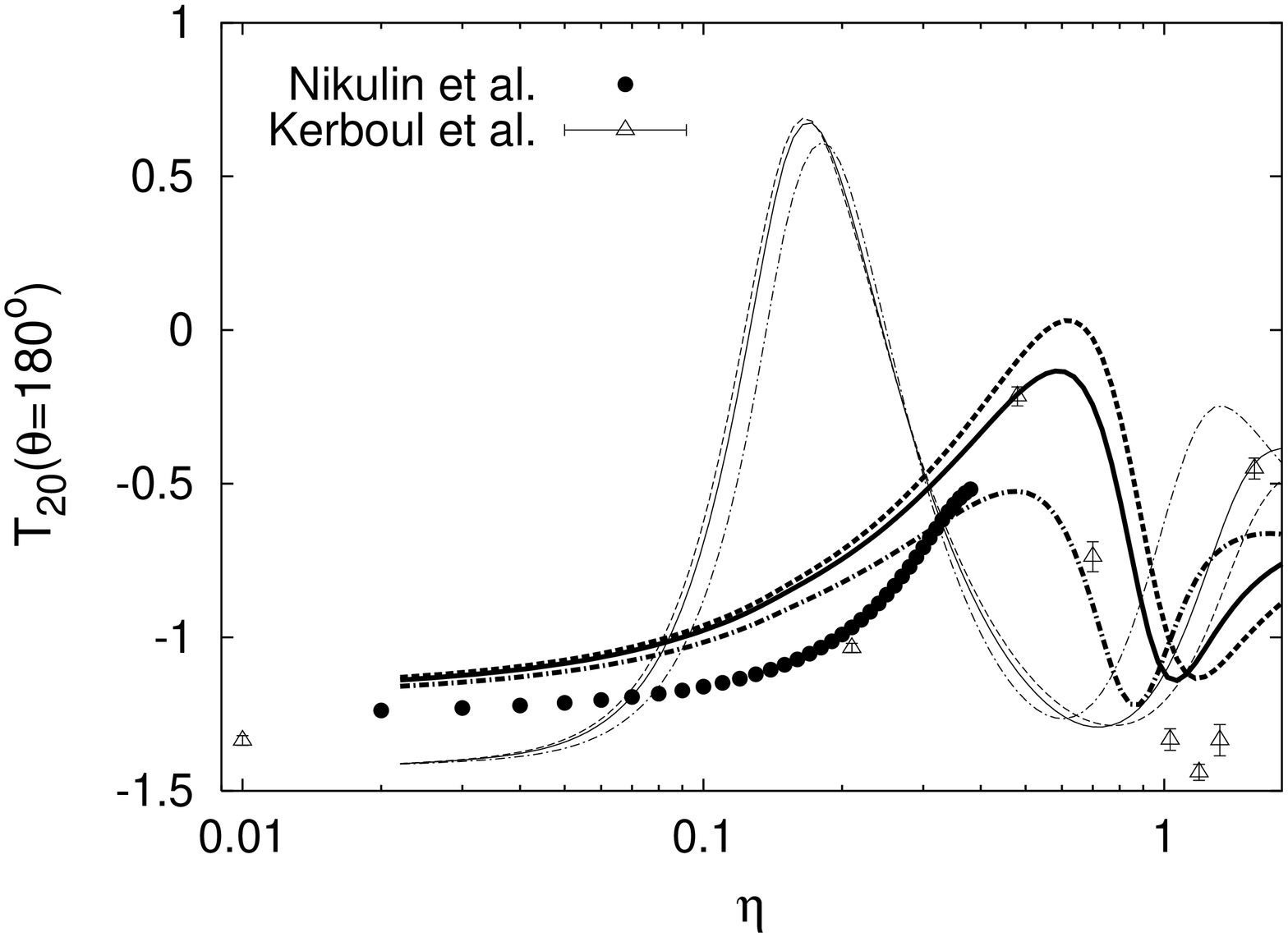, width=8cm, angle=-90}
\caption{Same as in Fig.4, but for backward direction.}
\end{figure}

In Fig. 1, the differential cross-section for the 
$dp\rightarrow {\ }^3He\ \pi^o$ process in collinear kinematics
is shown. In the considered energy range, the results in forward 
direction are always larger than the corresponding ones in backward direction, 
so that they are easily identified in the figure.
The theoretical curves are for bound-states (deuteron and 3N system)
calculated with different nucleon-nucleon potentials. Calculations of production
mechanisms include the isoscalar and isovector terms in S-wave, as well as the 
polar P-wave contribution and the $\Delta$-rescattering term.
The results indicate the sensitivity of the reaction to details of
the nuclear wavefunctions. Fig. 2 illustrates the effects on the same observable
caused by the various production mechanisms. The dot-dashed line includes
only the P-wave terms, while the dashed curve takes into account in addition
the effects of S-wave $\pi$-N rescattering in the isospin-odd 
(i.e., $\rho$-exchange) channel. The solid curve is the same of that in Fig. 1 
and depicts the results obtained when adding the rescattering term in the 
isospin-even channel. Similar results have been obtained also with
other nucleon-nucleon potentials.
The angular distribution of the cross section is shown in Fig. 3.
The results are compared with recent data obtained by the 
$GEM$ collaboration [15]. The corresponding energy range is in between 
the threshold region and the peak of the $\Delta$ resonance.
The data at backward angles agree with previous measurements performed at 
Saclay [13], but the new data in forward direction are substantially smaller.
The thin broken curve represents the theoretical calculations without
$\pi$N rescattering in the isospin-even (the so-called 
``$\sigma$''-exchange) channel; one obtains an evident underestimation
(by a factor ranging between two and three) of the differential cross-section in the 
forward direction. Inclusion of the r.h.s. term in Eq.~(\ref{s-waves-off-shell})
provides a much better description of the angular distribution, 
as shown by the two thick lines. The last two curves differ because
neutron-deuteron ISI have been included in the solid line 
via the solution of an AGS equation. To include ISI effects,
the nucleon-nucleon interactions in partial waves with 
$J\leq 2$ have been replaced with PEST-1 potentials\footnote{J.Haidenbauer, 
Private Communication}. 
Results suggest that ISI in the considered energy range do not play a major
role for spin-averaged observables. In a previous study [10] we found that
these ISI effects play a significant role in spin observables
involving strong interference effects amongst the various 
helicity amplitudes. This is the case of $A_y$ around the threshold region.

In Figs. 4 and 5 the deuteron tensor analyzing power $T_{20}$ is 
considered in collinear kinematics. In both directions, the $\pi N$ 
rescattering mechanism in the isoscalar channel has a major effect
on this observable, and is determinant for understanding the energy
dependence of $T_{20}$. It is worth observing that, 
very close to threshold, the results with the 
$\rho$-exchange diagram alone (thin lines) are close to $-\sqrt{2}$.
This is the geometrical limit for $T_{20}$ and is somewhat 
consistent with experiments performed in the threshold limit. However,
the energy dependence for this observable (as soon as one moves away
from the threshold limit) is totally out of phase, and one needs to include
also the term on the r.h.s. of Eq.~(\ref{s-waves-off-shell}) to
reproduce the trend of the experimental data. The results at backward
angles (Fig.5) exhibit also a significant dependence upon
the nucleon-nucleon potential employed in the calculation
of the nuclear bound-states. This suggests that in this way 
one could probe short-range aspects of the nucleon-nucleon correlations.

\vspace{4mm}
\noindent{\large \bf Summary and Conclusions}
\vspace{4mm}

We have calculated the $p+d\rightarrow \pi+(A=3)$ reaction employing
elementary production mechanisms obtained from the non-relativistic reduction
of the phenomenological pion-nucleon and pion-$\Delta$ Lagrangian of 
Eqs.~(\ref{Lag-piN},\ref{Lag-piD}). The corresponding matrix elements
between three-nucleon states have been evaluated within a 
large-basis space, consisting of 464 three-nucleon partial waves.
Total angular momenta of the system up to $7/2$ for both parities
have been considered. The prescriptions required by the Pauli principle 
have been taken into account through the application of the
permutation formalism to the three-nucleon system. Finally,
these matrix elements have been folded with the nuclear wavefunctions to 
obtain the probability amplitudes, and full details
of the procedure can be found in Refs.~[3,8-10].

The results obtained suggest that, in addition to the $\Delta$-rescattering
mechanism and to the $\rho$-exchange rescattering process in S-wave, 
one has to include also an S-wave rescattering mechanism generated  
by the $\pi$N interaction in the isoscalar channel. 
This channel is weak in elastic pion-nucleon
scattering, but could be well enhanced in meson production processes because
the corresponding kinematics is quite off-shell. At the present stage it is
not clear if these phenomenological off-shell effects, proposed many years ago
in the Literature (see Refs. [11,16,17]),
mimic more fundamental effects, such as for instance heavy-meson exchanges
contributions. If this is the case, it is surprising that 
the form expressed by the r.h.s. of Eq.~(\ref{s-waves-off-shell}) is needed 
not only for the absolute normalization of the cross-section at threshold, but also
for the shape of the angular distributions recently measured by 
the $GEM$ collaboration at energies ranging between the threshold region and 
the $\Delta$-resonance peak. And last but not least, the same effect
is able to account for the energy dependence of $T_{20}$ in collinear 
geometry in the energy range close to threshold, while the results obtained
without this contribution fail badly in reproducing this energy trend.

It is concluded that further investigations are needed to clarify these 
aspects. On the theoretical side, accurate investigations with 
alternative phenomenological models for the $\pi$N amplitudes should 
be welcome; 
in addition the role of the heavy-meson exchange diagrams
should be investigated separately. The numerical accuracy in treating 
the three-nucleon dynamics should be improved further, and the
role of relativity in the three-nucleon system for processes
above the pion threshold should be deeply investigated.
On the experimental side, further measurements are needed in order 
to constrain the ambiguities in the determination of the production 
amplitudes. With respect to this point, experiments to measure additional spin 
observables, such as $C_{yy}$ [18] in both collinear and 
non-collinear kinematics, should be considered of great utility.



\newcommand{\etal}{{\em et al. }}
\setlength{\parindent}{0mm}
\vspace{5mm}
{\bf References}
\begin{list}{}{\setlength{\topsep}{0mm}\setlength{\itemsep}{0mm}%
\setlength{\parsep}{0mm}}
%
\item[1.] M. Ruderman, Phys. Rev. {\bf 87}, 383 (1952).
\item[2.] W.R. Falk, Phys.Rev. C {\bf 61} 034005 (2000). 
\item[3.] L. Canton and W. Schadow, Phys. Rev. C {\bf 61}, 064009 (2000).
\item[4.] A.M. Green and E. Maqueda, Nucl. Phys. {\bf A316}, 215 (1979).
\item[5.] A.M. Green and M.E. Sainio, Nucl. Phys. {\bf A329} 477 (1979).
\item[6.] M.E. Sainio, Nucl. Phys. {\bf A389} 573 (1982).
\item[7.] E.O. Alt, P. Grassberger, W. Sandhas, Nucl. Phys. {\bf B2},
167 (1967).
\item[8.] L. Canton and W. Schadow, Phys. Rev. C {\bf 56}, 1231 (1997).
\item[9.] L. Canton \etal  Phys. Rev. C {\bf 57}, 1588 (1998).
\item[10.] L. Canton \etal Nucl. Phys. {\bf A684}, 417c (2001).
\item[11.] J. Hamilton, {\it High Energy Physics}, E.H.S. Burhop ed. 
(Academic, New York, 1967).
\item[12.] S. Schneider \etal {\em nucl-th/0209051} (from arXiv.org)  
\item[13.] C. Kerboul \etal $\ $ Phys. Lett. B {\bf 181}, 28 (1986).
\item[14.] V. Nikulin \etal $\ $ Phys. Rev. C {\bf 54}, 1732 (1996).
\item[15.] GEM {\em collaboration}, Nucl. Phys. {\bf A690}, 473 (2001).
\item[16.] O.V. Maxwell, W. Weise, and M. Brack, Nucl. Phys. {\bf A348}, 338 (1980).
\item[17.] E. Hernandez and E. Oset, Phys. Lett. {\bf B350} 158, (1995).  
\item[18.] V.P. Ladygin and N.B. Ladygina, Phys. Atom. Nucl. {\bf 58}, 1283 (1995). 
%
\end{list}

\end{document}